\documentclass[prl,aps,preprinnumbers,twocolumn,superscriptaddress,nofootinbib,balancelastpage]{revtex4}
\usepackage{amsthm}   
\usepackage{amsmath}  
\usepackage{amssymb}  
\usepackage{mathrsfs}
\usepackage[all]{xy}
\usepackage[pdftex]{graphicx}
\usepackage{indentfirst}
\usepackage[pdftex]{hyperref}

\newcommand{\be}{\begin{equation}}
\newcommand{\ee}{\end{equation}}
\newcommand{\ba}{\begin{eqnarray}}
\newcommand{\ea}{\end{eqnarray}}

\newcommand{\de}{\delta}

\newcommand{\sigmaG}{\sigma_{\mbox{\tiny G}}}

\begin{document}
\title{Protohalo Constraints to the Resonant Annihilation of Dark Matter}
\author{Francis-Yan Cyr-Racine}
\email{francis@phas.ubc.ca}
\affiliation{Department of Physics and Astronomy, University of British Columbia, Vancouver, BC, V6T 1Z1, Canada}
\author{Stefano Profumo}
\email{profumo@scipp.ucsc.edu}
\affiliation{%
Santa Cruz Institute for Particle Physics and Department of Physics,\\ University of California, Santa Cruz CA 95064, USA}
\author{Kris Sigurdson}
\email{krs@phas.ubc.ca}
\affiliation{Department of Physics and Astronomy, University of British Columbia, Vancouver, BC, V6T 1Z1, Canada}
\date{\today}
%
%
\begin{abstract}
It has recently been argued that the PAMELA, ATIC and PPB-BETS data showing an anomalous excess of high-energy cosmic ray positrons and electrons might be explained by dark matter annihilating in the Galactic halo with a cross section resonantly enhanced compared to its value in the primeval  plasma. We find that with a very large annihilation cross section the flash of energetic photons and electron-positron pairs expected from dark-matter annihilation in the first protohalos that form at redshift $z\sim40$ is likely substantial and observable.  As a consequence,  bounds on the allowed energy injection into the primordial gas and the energy density of the diffuse gamma-ray background give rise to limits on the low-velocity dark matter cross section that can be difficult to reconcile with this interpretation of the PAMELA, ATIC and PPB-BETS results.  
\end{abstract}
\maketitle
Recent data reported by several experiments may suggest the existence of a new source of cosmic ray positrons. Indeed, PAMELA \cite{Adriani:2008zr} has reported an excess in the positron fraction from 10 to 100 GeV while the ATIC \cite{:2008zzr} and PPB-BETS \cite{Torii:2008xu} experiments have detected an overabundance of charged leptons in the total positron-electron ($e^+e^-$) energy spectrum between 300 and 800 GeV (see also the recent Fermi LAT results \cite{abdo:181101}). A very interesting explanation of these data invokes the annihilation of dark matter particles with a mass at the TeV scale in the Galactic halo. However, in conventional models the dark matter annihilation cross section needed to account for the  excess is much larger (by a factor \mbox{$\mathcal{O}(100$\,to\,$1000)$}) than the value deduced from the observed dark matter relic abundance $\Omega_d h^2 \simeq 0.11$. To account for this, Refs.~\cite{Feldman:2008xs,Ibe:2008ye} (see also Ref.~\cite{Guo:2009aj}) propose that this ``enhancement factor" can be explained by a  resonance in the dark matter annihilation cross section (see Ref.~\cite{Ibe:2009dx} for an explicit realization).

 In this \emph{Letter}, we calculate the number and spectrum of photons and $e^+e^-$ pairs produced by the annihilation of dark matter to standard model (SM) particles in the first protohalos that form at redshift $z\lesssim40$.  We find that experimental constraints from the diffuse gamma-ray background and on the amount of energy injection allowed into the primordial medium can be difficult to reconcile with the large annihilation cross sections \mbox{$\sigma \!\sim\!10^{-6}$ to $10^{-7}$ GeV$^{-2}$} required to account for the observed Galactic lepton excess.

While we focus here on models with a Breit-Wigner resonance in the dark matter annihilation cross section our constraints to $\sigma_0$, the low velocity-dispersion annihilation cross section, are model-independent and apply to any model in which dark matter annihilates predominately to SM final states (for instance Ref.~\cite{ArkaniHamed:2008qn}). We begin by briefly reviewing the resonant enhancement mechanism before deriving the constraints on the cross section from protohalo collapse. We then use current experimental bounds from diffuse backgrounds and energy injection into the primordial gas to constrain the parameter space of the resonant cross section.
\newline
 
\begin{figure}[t] 
   \centering
   \includegraphics[width=0.49\textwidth]{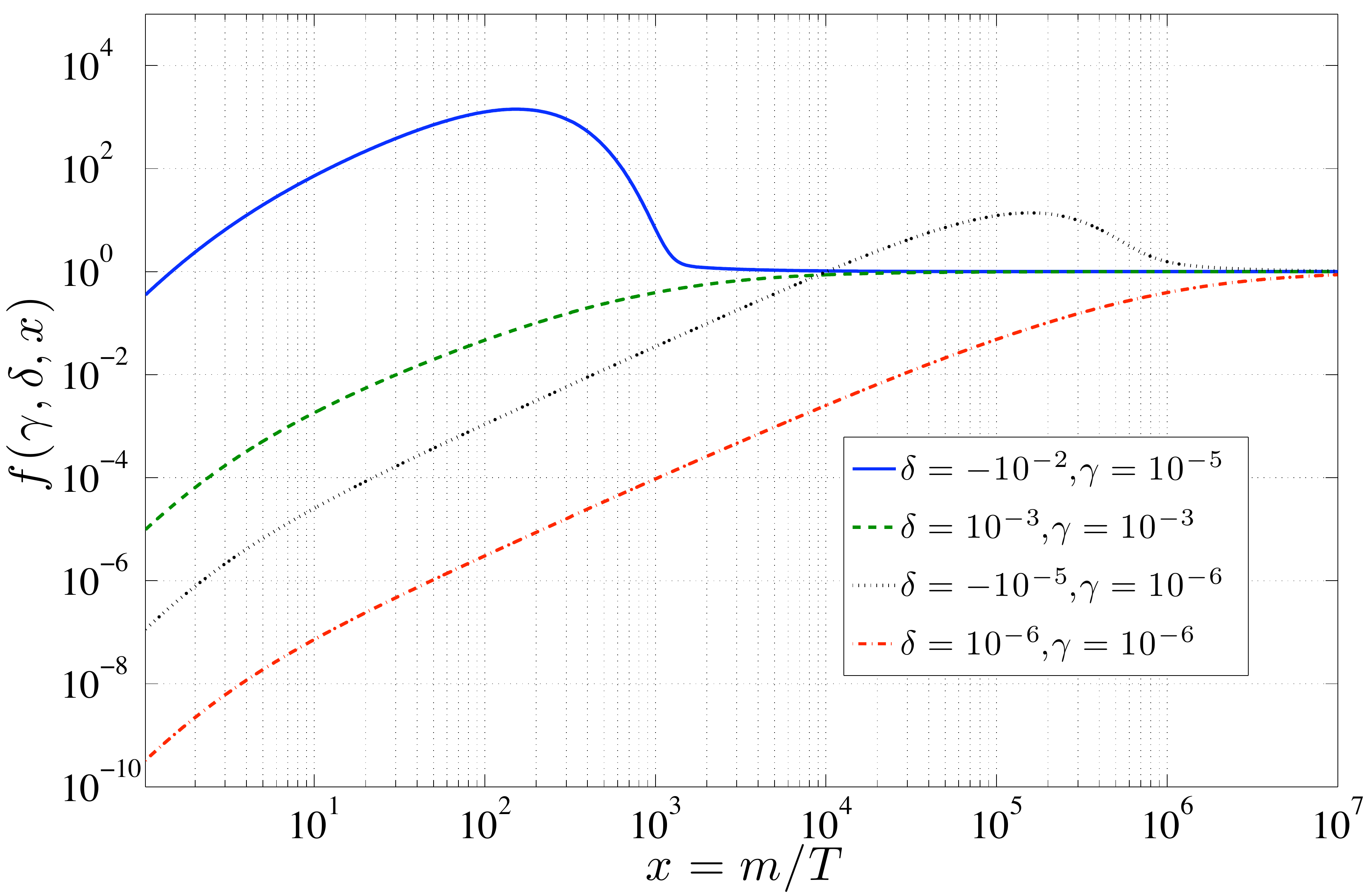} 
   \caption{The functional form of the resonance function $f(\gamma,\delta,x)$ for various $\delta$ and $\gamma$  with $g_i=2$.}
   \label{figure_xsection}
\end{figure}

\noindent\emph{Breit-Wigner Resonance} -- We consider a model in which two dark matter particles of mass $m$ and energy $E_{i=1,2}$ annihilate via a narrow resonance of mass $M$. Following Refs.~\cite{Ibe:2008ye,Guo:2009aj}, we parametrize this resonance using
\be\label{def_delta}
M^2=4m^2(1-\delta),\quad |\delta|\ll1.
\ee
For $\delta<0$, we have a physical pole (particle state) while for $\delta>0$, we have an unphysical pole. In both cases, the cross section times velocity takes the form
\be
\hat{\sigma}(z)\equiv4E_1E_2\sigma v\propto\frac{(1+z)\gamma^2}{(z+\delta)^2+\gamma^2},\quad\gamma\equiv\Gamma/M
\ee
where the Mandelstam variable $s\!\!=\!\!4m^2(1+z)$ and $\Gamma$ is the decay width of the resonance. To calculate the relic abundance of dark matter, we thermally average the annihilation cross section $\sigma$
\be\label{thermal_average}
\langle\sigma v\rangle=\frac{g_i^2}{n_{EQ}^2}\frac{m^4}{8\pi^4x}\int_{0}^{\infty}\!\!\!dz\,\sqrt{z}\,\hat{\sigma}(z)K_1\left(2x\sqrt{1+z}\right)\,,
\ee
where $x\equiv m/T$ and $n_{EQ}=(g_i m^3/2\pi^2)K_2(x)/x$.
Here $K_1(x)$ and $K_2(x)$ are modified Bessel functions and $g_i$ is the number of helicity states of a dark matter particle. Evaluating the integral (\ref{thermal_average}), we can write the thermal cross section as $\langle\sigma v\rangle=\sigma_0f(\delta,\gamma,x)$, where the function $f$ encodes all the information about the resonance and has the property $f(\delta,\gamma, x\gg1)=1$. While there is no simple analytic expression for $f$, it can straightforwardly be found numerically (see Figure~\ref{figure_xsection}). To determine the relic density, we solve the Boltzmann equation
\be\label{boltzmann}
\frac{dY}{dx}=-\frac{\lambda}{x^2}f(\delta,\gamma,x)(Y^2-Y_{EQ}^2)
\ee
for the dark matter yield, $Y=n/s$, where $n$ is the number density of dark matter and $s$ is the entropy density. Here, $\lambda=\sqrt{8\pi^2g_*/45}M_{pl}m\sigma_0$ where $g_*$ is the number of relativistic degrees of freedom, $M_{pl}$ is the reduced Planck mass and
\be
Y_{EQ}=\frac{45}{4\sqrt{2}\pi^{7/2}}\left(\frac{g_i}{g_*}\right)x^{3/2}e^{-x}\nonumber
\ee
is the equilibrium dark matter yield. In the usual nonresonant scenario with $f \rightarrow 1$, $Y$ tracks $Y_{EQ}$ until the annihilation rate falls below the Hubble expansion rate and the interactions freeze out. The freeze-out temperature $T_f=m/x_f$ is conventionally determined when $Y-Y_{EQ}\simeq\mathcal{O}(Y_{EQ})$ --- when the yield $Y$ deviates substantially from equilibrium.  The relic abundance is then given by the solution for $Y$ at late time, $Y_{\infty}\simeq x_f/\lambda$.

However, in the presence of a resonance, the annihilation process does not freeze out when $Y-Y_{EQ}\simeq\mathcal{O}(Y_{EQ})$ as $f(\delta,\gamma,x)$ is increasing with $x$ (for all $\de>0$ and some cases with $\de<0$). Instead, the dark matter keeps annihilating until a much lower temperature $T_b=m/x_b\ll T_f$ and the relic abundance is given by the asymptotic solution $Y_{\infty}^{res}\simeq x_b/\lambda$. The resonant case must have a higher $\lambda$ (and thus $\sigma_0$) to obtain a relic abundance compatible with cosmological observations. In particular, if this higher $\sigma_0$ leads to an important production of positron-electron pairs in the Galactic halo, then one could explain the anomalous leptonic excess as was argued in Ref.~\cite{Ibe:2008ye}.
\newline

\noindent\emph{Annihilation in protohalos} -- After matter-radiation equality, perturbations in the dark matter start to grow via gravitational instability and form protohalos --- the first nonlinear structures in the Universe.  The formation of protohalos  triggers a flash of annihilation of dark matter particles  at redshift $z_f \sim40$ \cite{Kamionkowski:2008gj,Oda:2005nv}. As they have a small mass and a low velocity dispersion \cite{kris}, the annihilation cross section in these protohalos is given by $\sigma_0$. 
If dark matter annihilates into SM particles, a significant fraction of the initial energy will be converted to high-energy photons and $e^+e^-$ pairs.  A photon generated can either travel freely through the Universe if its energy is within the gap between 0.1 MeV$\lesssim E_{\gamma}\lesssim$ 0.3 TeV in which the Universe is essentially transparent \cite{Chen:2003gz}, or it is absorbed by the primordial gas. The $e^+e^-$ pairs produced rapidly inverse-Compton scatter off CMB photons resulting in gamma rays that are similarly either absorbed by the primordial gas or freestream if their energies are in the transparent gap.  Experimental bounds on the diffuse extragalactic background of gamma rays and on the energy injection into the primordial gas constrains the number of $e^+e^-$ pairs and photons that could have been created in the first dark matter halos. As we now show, this bounds the dark matter cross section to annihilate into $e^+e^-$ pairs.
\newline

\noindent\emph{Constraints from the Diffuse Background} -- As the density inside a virialized protohalos is $\sim\!\!180$ times higher than the mean cosmological density at  redshift $z_f$, the annihilation rate $\Gamma=n\langle\sigma v\rangle$ in protohalos is
\be
\Gamma\simeq 4.9 \times10^{-6}\left(\frac{B\sigma_0}{\text{GeV}^{-2}}\right)\left(\frac{m}{\text{TeV}}\right)^{-1}\left(\frac{z_f}{40}\right)^{3}\text{Myr}^{-1}
\ee
assuming the present dark matter density to be \mbox{$\Omega_{d}h^2\simeq0.11$}. We have introduced the standard boost factor $B$ to account for the nonuniform distribution of dark matter in these halos. Plausible values for $B$ range between 3 and 60 depending on how the halos are concentrated \cite{Kamionkowski:2008vw}.

As the Universe expands nonlinear structures form via hierarchical collapse and the total fraction of dark matter particles bound in collapsed objects increases \cite{Cooray:2002dia}.  Protohalos eventually merge into more massive halos with a lower mean density and mean annihilation rate --- although the dense cores of first-generation halos likely continue to shine relatively brightly for some time as dense substructures in larger halos.

We find that the fraction $\Theta$ of dark matter particles that annihilate in protohalos and other dense structures is
\be\label{fraction}
\Theta\simeq3.9\times10^{-4}\left(\frac{B\sigma_0}{\text{GeV}^{-2}}\right)\left(\frac{m}{\text{TeV}}\right)^{-1}\left(\frac{z_f}{40}\right)^{3/2},
\ee
where $\Theta\simeq (1/3)\Gamma \Delta t|_{z_f\simeq40}$.  We model dense structures that collapse at redshift $z$ to annihilate efficiently for an expansion ($e$-folding) time $\Delta t$ before being disrupted so that annihilation shuts off. At redshift $z_f$, $\Delta t\sim2.4\times10^2(z_f/40)^{-3/2}$ Myr and the factor of $1/3$ accounts for the facts that: \!\!($i$) only a fraction of the Universe has collapsed into nonlinear structure at the redshifts of interest; ($ii$) structure forming at $z \lesssim z_f$ and $z \gtrsim z_f$ also contributes to the mean annhilation rate of the Universe.   We find Eq.~(\ref{fraction}) evaluated in the `flash approximation' at redshift $z_f$ is a good estimate for detailed calculations of the mean annihilation rate using Press-Schecter theory.

If the photons generated are not absorbed by the primordial gas, then they contribute to the diffuse background of gamma rays with energy density $\rho_{\gamma}=\Theta\rho_{{\rm crit}}\Omega_{d}/z_f$, where $\rho_{{\rm crit}}$ is the critical density of the Universe today and we have accounted for the redshift of the photons. Using Eq.~(\ref{fraction}), we find
\be
\rho_{\gamma}\simeq1.1\times10^{-11}\left(\frac{\Omega B\sigma_0}{\text{GeV}^{-2}}\right)\left(\frac{m}{\text{TeV}}\right)^{-1}\left(\frac{z_f}{40}\right)^{\frac{1}{2}}\frac{\text{GeV }}{\text{cm}^{3}},
\ee
where $\Omega$ is the fraction of the initial energy that is converted to photons (or electron-induced photons) whose energies lie inside the transparent gap. A fit from EGRET \cite{Sreekumar:1997un} to the gamma-ray spectrum of unresolved astrophysical sources  yields the bound $\rho_{\gamma}^{EGRET}\!\!\approx\!5.7\times10^{-16}(E_{\gamma}/\text{GeV})^{-0.1}$ GeV cm$^{-3}$. Assuming that this energy excess is entirely accounted for by annihilating dark matter in the first structures, we obtain
\be\label{constraint1}
\sigma_0\lesssim\frac{5.0\times10^{-5}}{B\Omega}\left(\frac{m}{\text{TeV}}\right)\left(\frac{z_f}{40}\right)^{-\frac{1}{2}}\left(\frac{E_{\gamma}}{\text{GeV}}\right)^{-0.1}\text{GeV}^{-2}.
\ee
This bound is a conservative upper limit on the annihilation cross section as other contributions to the gamma ray background are likely present. Forthcoming data from the Fermi experiment should improve this limit \cite{Meurer:2009ir}.
\newline

\noindent\emph{Constraints on Energy Injection into Primordial Gas} -- We now consider the case for which the energy released by the annihilating dark matter is absorbed by the primordial gas. Detailed modeling of CMB and large-scale-structure data \cite{Zhang:2007zzh} yield a bound $\Theta\lesssim3\times10^{-10}$ on the fraction of the total rest mass energy of dark matter that could have been injected in the primordial gas when the age of the Universe was $t_f \sim 67$ Myr. Using Eq.~(\ref{fraction}), we then find
\be\label{constraint2}
\sigma_0\lesssim\frac{7.6\times10^{-7}}{B\Omega'}\left(\frac{m}{\text{TeV}}\right)\left(\frac{z_f}{40}\right)^{-\frac{3}{2}}\text{GeV}^{-2},
\ee
where $\Omega'$ is the fraction of the initial energy that is injected in the form of photons whose energies lie above the transparent gap (i.e. photons with $E_{\gamma}\gtrsim300$ GeV either generated promptly or via inverse-Compton scattering).  Forthcoming results from the Planck satellite are likely to strengthen this bound.
\newline

\noindent\emph{Discussion} -- If the anomalous leptonic signal is accounted for by annihilating dark matter, the value of the cross section to $e^+e^-$ pairs in the Galaxy must be in the range $\sigma_{e^+e^-}\!\sim\!10^{-6}$ to $10^{-7}$ GeV$^{-2}$. The total annihilation cross section in the galaxy today, $\sigmaG$, is related to $\sigma_0$ by a transfer function $g(\delta,\gamma)\equiv\sigmaG/\sigma_0 \simeq f(\delta,\gamma,x_{\mbox{\tiny G}})$, where we take  $x_{\mbox{\tiny G}}\!\!\sim\!3\times10^{6}$ in the Galactic halo \cite{Guo:2009aj}. This function accounts for differences between $\sigma_0$ and $\sigmaG$ for very small $\gamma$ and $\delta$ ($\delta,\gamma\lesssim10^{-5}$). We find $g\!\!\sim\!\!0.67$ to $1$ for $10^{-4}\!\geq\!\de,\gamma\!\geq\!10^{-6}$ and $g\!\!\sim\!\!1$ to $1.65$ for $-10^{-6}\!\leq\!\de,\gamma\!\leq\!-10^{-4}$ while for $|\de|,\gamma\!\gtrsim\!10^{-4}$, we find $g(\de,\gamma)\!=\!1$. As the cross section to $e^+e^-$ pairs is necessarily smaller then the total annihilation cross section, we have $\sigma_{e^+e^-}\!\leq g(\delta,\gamma)\sigma_0$.

\begin{figure}[t]
   \centering
   \includegraphics[width=0.45\textwidth]{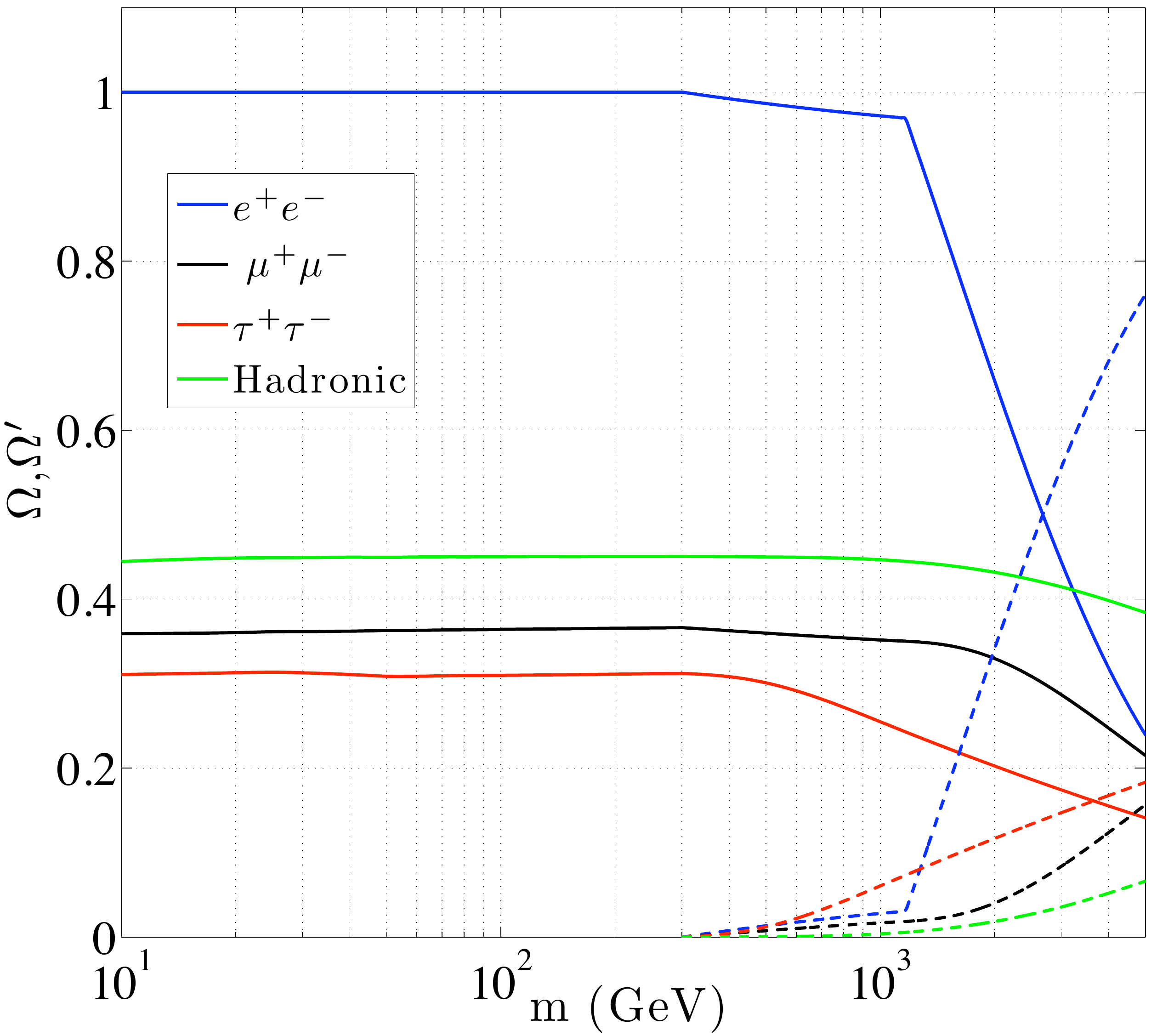} 
   \caption{Fraction of the initial energy (twice the dark matter mass) that is converted to photons and electron-induced photons with energies inside ($\Omega$, full lines) the transparent gap ($0.1$ MeV $\lesssim E_{\gamma}\lesssim 0.3$ TeV) as well as outside ($\Omega'$, dashed lines) the transparent gap ($E_{\gamma}\gtrsim 0.3$ TeV) as a function of the dark matter mass for different annihilation channels. The hadronic case assumes equal probability of annihilating to any of the quarks or to gluon pairs.}
   \label{omega}
\end{figure}

To calculate the fraction of the total initial energy that is converted to photons (and electron-induced photons) inside ($\Omega$) and outside ($\Omega'$) the transparent energy gap, we use Monte Carlo simulations of the photon and  $e^+e^-$ pairs spectra and yields obtained from \emph{DarkSUSY} \cite{Gondolo:2004sc}.  To accurately determine the energy injected via electron-induced photons we use the exact photon distribution expected from high energy inverse-Compton scattering with a Klein-Nishina (KN) cross section (see Appendix A of Ref.~\cite{1990MNRAS.245..453C}).  This is important because for electron energies $E_e \sim\,$TeV typical CMB photons at $z \sim 40$ have energies comparable to $m_e$ in the electron rest frame and KN corrections are significant.

We consider four fiducial cases in which the dark matter annihilates either only into $\tau^+\tau^-$, $\mu^+\mu^-$ or $e^+e^-$ pairs, or only into hadrons, with equal probability of annihilating into any of the $q$-$\bar{q}$  pairs or to a gluon pair. For the hadronic case, the main contribution comes from photons produced promptly by dark matter annihilation while electron-induced photons contribute at most $\sim40$ percent to the total energy. On the other hand, electron-induced photons contribute most of the energy fraction for the muon case while the tau channel is dominated by direct photon production.
\begin{figure}[b]
   \centering
   \includegraphics[width=0.45\textwidth]{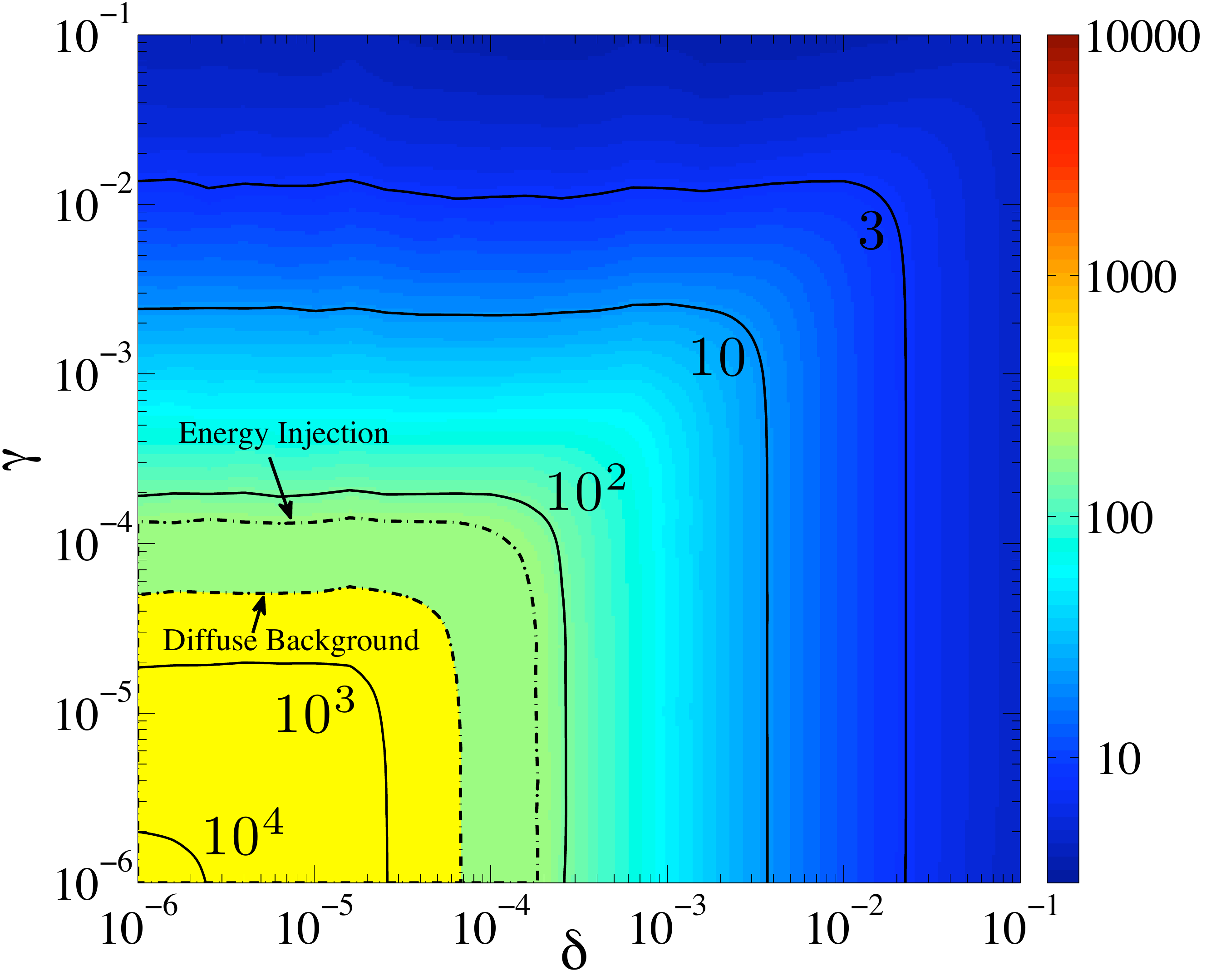} 
   \includegraphics[width=0.45\textwidth]{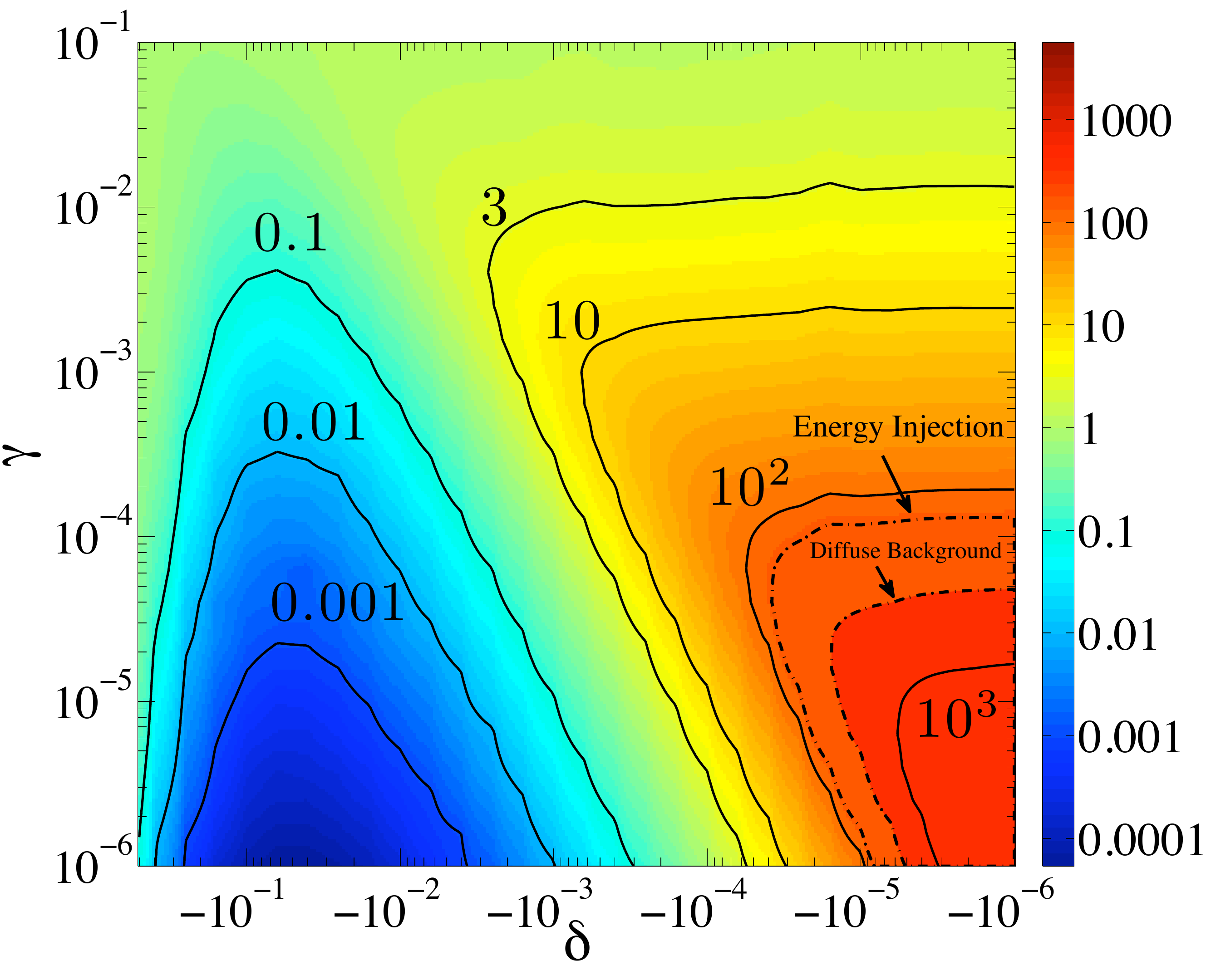} 
   \caption{Value of the Enhancement Factor ($\equiv Y_{\infty}^{res}/Y_{\infty}$) of the dark matter annihilation cross section for an unphysical pole (top) and a physical pole (bottom) for models that yield $\Omega_d h^2\simeq0.11$. The dash-dot lines delimit regions where values of $\delta$ and $\gamma$  are excluded by (\ref{const_final1}) and (\ref{const_final2}) as indicated. Note that the shading in these excluded regions reflects the value on the constraint boundary.}
   \label{figure_constraints}
\end{figure}

In Figure \ref{omega}, we plot the energy fraction $\Omega$ (full lines) as a function of the dark matter mass for the four channels. A realistic model might include a mixture of hadronic and leptonic annihilations (although current Galactic data may favor a leptophilic process) which would lead to an energy fraction $0.31\lesssim\Omega\lesssim1$ for $m=200$ GeV. Using the constraint Eq. (\ref{constraint1}) and taking $B\!\sim\!35$, $m=200$ GeV, $z_f=40$ and $\Omega\sim0.6$, we obtain
\be\label{const_final1}
\sigma_{0}\lesssim3.8\times10^{-7}\text{GeV}^{-2},
\ee
where we take $E_{\gamma}\sim10$ GeV, the energy with the highest flux for $m=200$ GeV. This constraint is shown in Figure \ref{figure_constraints} by the dash-dot line labelled ``Diffuse Background". For the allowed values of $\de$ and $\gamma$, $g(\de,\gamma)$ is between 0.97 and 1.3 and therefore the constraint on the cross section to $e^+e^-$ pairs is $\sigma_{e^+e^-}\!\!\lesssim(3.7-4.9)\times10^{-7}\text{GeV}^{-2}$. This bound excludes the resonant enhancement mechanism as a solution for the positron fraction excess problem for some of plausible range for $m$ and $B$ (although for a low enough value of $B/m$ a solution might still be found).
\begin{figure}[t]
   \centering
   \includegraphics[width=0.45\textwidth]{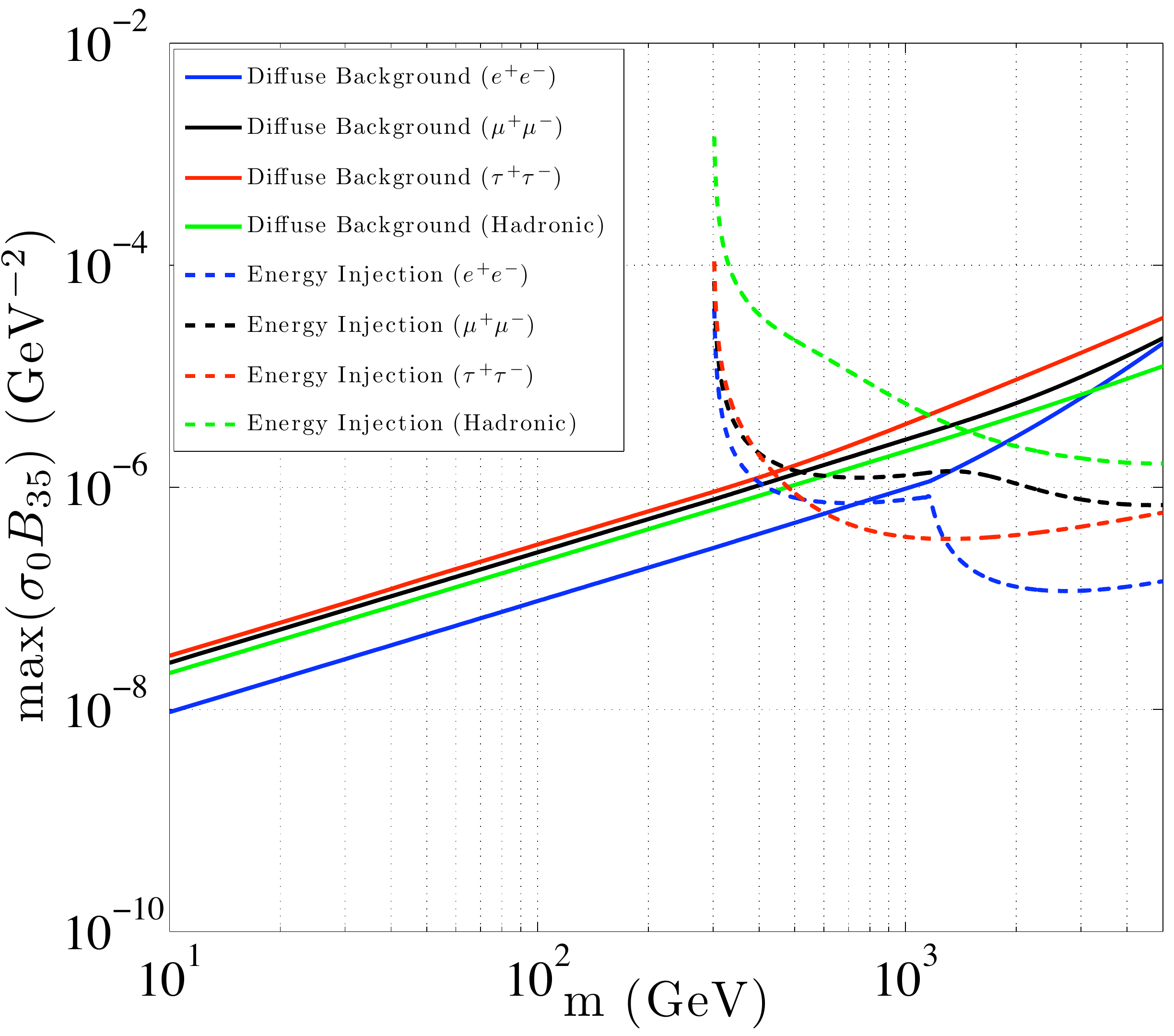} 
   \caption{Upper bound on $\sigma_0B_{35}$ from (\ref{const_final1}) and (\ref{const_final2}) for both leptonic and hadronic annihilation of dark matter ($z_f=40$). The regions above the curves are excluded. ($B_{35}\equiv B/35$)}
   \label{model_ind_const}
\end{figure}

For $m\gtrsim600$ GeV, a stronger bound can be put on the cross section to positron-electron pairs using Eq.~(\ref{constraint2}). In Figure \ref{omega}, the dashed lines show the energy fractions above the transparency window $\Omega'$ as a function of $m$ for the three leptonic cases and the hadronic case. Again, a realistic model might involve some mixture of the two and therefore $0.01\lesssim\!\Omega'\!\lesssim0.34$ at $m=2$ TeV. With $\Omega'\sim0.3$, $z_f=40$ and $B\sim35$ in Eq. (\ref{constraint2}), we find
\be\label{const_final2}
\sigma_0\lesssim1.4\times10^{-7} \text{GeV}^{-2}.
\ee
This constraint is shown in Figure \ref{figure_constraints} by the dash-dot lines labelled ``Energy Injection". Eq. (\ref{const_final2}) translates directly to a bound on $\sigma_{e^+e^-}$. Such a cross section is not large enough to account for the $e^+e^-$ excess observed by the satellite experiments. One could weaken this constraint by allowing for a smaller value of $B/m$.

Finally, generalizing Eqs.~(\ref{const_final1}) and (\ref{const_final2}) gives the model-independent upper bounds on $\sigma_0B_{35}$ shown in Figure \ref{model_ind_const} ($B_{35}\!\equiv\!B/35$). We see that light dark matter ($m\!\!<\!\!100$ GeV) is excluded by the diffuse background constraint if the anomalous leptonic signal is to be explained by dark matter annihilating in the Galactic halo. The energy injection constraints  for charged lepton-pair production disfavor a dark matter mass at the TeV scale.
\newline

\noindent\emph{Summary} -- We have shown that a resonant dark matter annihilation cross section to $e^+e^-$ pairs large enough to explain the Galactic lepton anomalies is in tension with data from the diffuse gamma ray background and limits on energy injection into primordial gas.   The high enhancement regions of the parameter space are difficult to reconcile with these bounds assuming that protohalos are not exceptionally diffuse.
Forthcoming data from the Fermi satellite might detect telltale signatures of dark matter annihilation or yield even more stringent constraints to resonant annihilation models.

\acknowledgments
F-Y.C-R. is supported by the Natural Sciences and Engineering Research Council (NSERC) of Canada.  S.P. is partly supported by US DoE Contract DEFG02-04ER41268 and by NSF Grant PHY-0757911.   K.S. is supported by a NSERC Discovery Grant.
\bibliography{resub_v1}
\end{document}